\documentclass[amssymb,11pt]{article}
\usepackage[pdftex,colorlinks=true,urlcolor=black,filecolor=black,linkcolor=black,
            pdftitle={A Characteristic Particle Length.},
            pdfauthor={Mark D. Roberts},
            pdfsubject={Schrodinger equation},
            pdfkeywords={lenght,distance,time},
            pagebackref,pdfpagemode=None,bookmarksopen=true]{hyperref}
\usepackage{amssymb}
\usepackage{graphicx}
\newcommand{\bc}{\begin{center}}
\newcommand{\ec}{\end{center}}
\newcommand{\be}{\begin{equation}}
\newcommand{\ee}{\end{equation}}
\newcommand{\ber}{\begin{eqnarray}}
\newcommand{\ear}{\end{eqnarray}}

\begin{document}
\title{A Characteristic Particle Length.}
\author{
\href{http://www.violinist.com/directory/bio.cfm?member=robemark}
{Mark D. Roberts},\\
}
\date{$14^{th}$ of June 2014}
\maketitle
\begin{abstract}
It is argued that there are characteristic intervals associated with any particle that can
be derived without reference to the speed of light $c$.  Such intervals are inferred from
zeros of wavefunctions which are solutions to the Schr\"odinger equation.  The characteristic
lenght is $\ell=\beta^2\hbar^2/(8Gm^3)$,  where $\beta=3.8\dots$;  this lenght might lead to
obsevational effects on objects the size of a virus.
\end{abstract}
{\tableofcontents}
\section{Introduction.}\label{intro}
Consider the spreading of the wave-packet in non-relativistic quantum mechanics.
One can ask at what point does gravitational attraction stop its spreading,
compare \cite{roberts} where it was asked whether the deviation of geodesics
and wave spreading could cancel;
however as it stands the wave-packet is a solution to Schr\"odinger's equation
\cite{schiff}eq.6.16
\begin{equation}
\label{schroeq}
i\hbar\frac{\partial\psi}{\partial t}=-\frac{\hbar^2}{2m}\nabla\psi+V\psi,
\end{equation}
and so obeys Ehrenfest's theorem
\cite{schiff}eq.7.10
\begin{equation}
\label{ehrenfest}
\frac{d}{d t}<p_a>=-\left<V_{,a}\right>,
\end{equation}
but the potential $V$ was taken to vanish in the derivation of the wavefunction so that
gravitational attraction cannot be taken into account.   To overcome this one has to start
again and derive the wavefunction with a non-vanishing gravitational
potential from the beginning .

Non-relativistic quantum mechanics and newtonian gravity are also used to describe
the cow experiment \cite{cow}.
Here throughout charge and spin are ignored although for most particles this is a big
assumption,  this is done for reasons of simplicity.
The conventions used are:  a wavefunction is any solution to the Schr\"odinger equation,
a wave-packet is a wavefuntion with a distribution which has a well-defined mean and variance,
signature $-,+,+,+$.
The use of $d$ is avoided as it can both denote distance and dimension,  $n$ is used to denote
the number of spatial dimensions.   It turns out that $n=1,2,4$ are all exceptional cases,
so to avoid equation clutter we usually stick to $d=3$ unless stated otherwise.
Some constants,  see \cite{barrow},  used are:
first zero of the Bessel J function $\beta=3.831705970$,
Euler's constant $\gamma=0.5772156649$,
Planck's constant divided by $2\pi$ $\hbar=1.054571726(47)\times10^{-34}Js$,
gravitational constant $G=6.67384(80)\times10^{-11}m^3kg^{-1}s^{-2}$,
speed of light $c=2.99792458\times10^8ms^{-1}$.
The Planck units are
\begin{equation}
\label{planckunits}
m_p\equiv\sqrt\frac{\hbar c}{G},~~~
l_p\equiv\sqrt\frac{\hbar G}{c^3},~~~
t_p\equiv\sqrt\frac{\hbar G}{c^5}.
\end{equation}
\section{Static Case.}\label{staticcase}
Taking the time independent Schr\"odinger equation,  (\ref{schroeq}) with vanishing LHS,
and with vanishing potential $V=0$,  the spherically symmetric solution is
\begin{equation}
\label{2.1}
\psi(r)=A+\frac{B}{r},
\end{equation}
where $A$ and $B$ are amplitude constants.  Now taking the Newtonian gravitational potential
\begin{equation}
\label{newtpot}
V=-\frac{GM{\tt m}}{r},
\end{equation}
equating the Schr\"odinger mass $m$,  the gravitating mass $M$ and the test mass ${\tt m}$
and using the notation
\begin{equation}
\label{defk}
k\equiv \frac{2Gm^3}{\hbar^2},
\end{equation}
where $k$ is of dimensions $L^{-1}$,  the time independent spherically symmetric Schr\"odinger
equations becomes
\begin{equation}
\label{besseq}
0=\psi_{rr}+\frac{2}{r}\psi_r+\frac{k}{r}\psi,
\end{equation}
which is a Bessel equation with solution
\begin{equation}
\label{bessol}
\psi(r)=\frac{C}{\surd r}BesselJ(1,2\sqrt{kr})+\frac{D}{\surd r}BesselY(1,2\sqrt{kr}),
\end{equation}
where $C$ and $D$ are amplitude constants,  see the figure.
\begin{figure}
\includegraphics[height=6.0in]{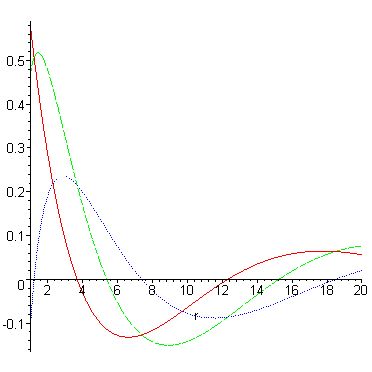}
\caption{
$BesselJ(1,2\sqrt(r))/\sqrt(r)$,
$BesselY(1,2\sqrt(r))/\sqrt(r)$,
$BesselJ(1,2\sqrt(r))/\sqrt(r)+BesselY(1,2\sqrt(r))/\sqrt(r)$}
\end{figure}
Note $\beta^2/4=3.67049266$ is where the intercept is.
Briefly for $n=1$ (\ref{bessol}) is replaced by trigonometric functions,
for $n=2$ (\ref{2.1}) and (\ref{newtpot}) involve logs,
for $n=4$ the Bessel order and argument diverge,
for $n\ge5$ the external $\surd r\rightarrow r^{1-n/2}$,
the argument $\surd r\rightarrow r^{2-n/2}$
and the order $1\rightarrow (n-2)/(4-n)$.
Expanding (\ref{bessol}) to first order in $r$ and choosing no mixing of the terms
\begin{equation}
\label{newcons}
A=\sqrt{r}C,~~~B=-\frac{D}{\pi\sqrt{k}},
\end{equation}
gives the lowest order correction to (\ref{2.1})
\begin{equation}
\label{lowest}
\psi(r)=A\left[1-\frac{kr}{2}+\dots\right]
+\frac{B}{r}\left[1+kr(1-2\gamma)-kr\ln(\sqrt{kr})+\dots\right].
\end{equation}
At first sight this is counter intuitive as one would expect the addition of a potential
to add short range decaying terms to the wavefunction;  however one should think of the
Schr\"odinger equation as a statement of the conservation of energy and gravitational
energy is negative hence the increasing terms.
That (\ref{bessol}) sometimes has negative wavefunction is not necessarily unphysical
as it is products $\psi\psi^*$ that correspond to measurable quantities.
There is the question of what the zeros of $\psi$ correspond to.
The solutions (\ref{2.1}) and (\ref{bessol}) are pre-interpretation solutions to the
Schr\"odinger equation in the sense that one cannot construct expectations to the momenta
and so forth as there is no time-dependence or overall energy:  pre-interpretational can
be thought of as the Schr\"odinger equation (\ref{schroeq}) with the LHS taken to vanish.
Another way of thinking of this is that a solution (\ref{2.1}) or (\ref{bessol}) is a choice
of vacuum,  so that normally one choose only $A\ne0$ but when self-gravitation is taken into
account the simplest choice is only $C\ne0$.   Once this choice has been taken one has a
critical distance where the wavefunction vanishes
\begin{equation}
\label{critlen}
\ell\equiv r_{crit}=\frac{\beta^2}{4k}=\frac{\beta^2\hbar^2}{8Gm^3}.
\end{equation}
\section{Non-static Case.}\label{nonstatic}
\begin{figure}
\includegraphics[height=2.3in]{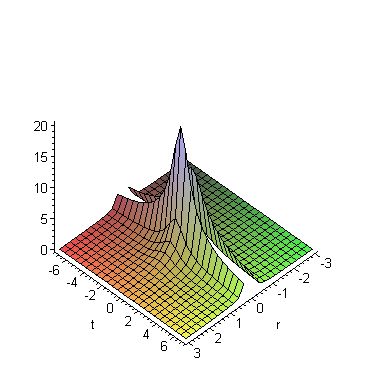}
\caption{Matterhorn of A and B amplitudes both non-vanishing}
\end{figure}
In the non-static case there is the time dependent Gaussian wave-packet solution is
\begin{equation}
\label{nstaticsol}
\psi(r,t)=\left[A+B\left(\frac{r}{f(t)}\right)^{(2-n)}\right]f(t)^{-\frac{n}{2}}\exp\left(-\frac{r^2}{2f(t)\sigma^2}\right)
\end{equation}
where
$f(t)\equiv 1+i\hbar t/(m\sigma^2)$,
$A,~B$ are constants as in (\ref{2.1}),
$\sigma$ is the raw variance.
The $A$ term corresponds to a flat solution with a Gaussian added,
the $B$ term corresponds to a reciprocal point particle potential with a Gaussian added
this term diverges as $r$ goes to $0$,  the terms can be added as can be checked explicitly
and as would be anticipated from the superposition principle.
The solution is to a $V=0$ Schr\"odinger equation, adding $V=-\alpha/r^2$ is straightforward,
time dependent solutions for other potentials in particular for $V=-k/r$ are unknown.
The solution can be expressed in terms of modified Whittaker functions
\begin{equation}
\label{whitsol}
\psi(r,t)=\left[A+B\left(\frac{r}{f(t)}\right)^{(2-n)}\right]f(t)^{-\frac{n}{2}}
\left(-\frac{r^2}{\sigma^2f(t)}\right)^\alpha {\rm WhittakerM}\{\alpha,-\alpha-1/2,-\frac{r^2}{\sigma^2f(t)}\}
\end{equation}
or hypergeometric functions
\begin{equation}
\label{hypsol}
\psi(r,t)=\left[A+B\left(\frac{r}{f(t)}\right)^{(2-n)}\right]f(t)^{-\frac{n}{2}}
{\rm hypergeom}\{[-2\alpha],[-2\alpha],-\frac{r^2}{2\sigma^2f(t)}\},
\end{equation}
where $\alpha$ is a constant.
There does not appear to be a solution with the Gaussian distribution replaced by any
Pearson distribution or a similar distribution,
however because to the radially dependent terms in front
of the gaussian there is now non-vanishing excess kurtosis,
and $\sigma$ is no longer the variance which is why it is called the raw variance above.
Setting $n=3,~~~A=B=m=\hbar=\sigma=1$ then $\psi\psi^*$
gives the Matterhorn shape in the figure,
there is a divergence at the origin as would be expected
for $B\ne 0$ because of the divergence of the reciprocal potential.

There seems to be no time dependent version of the Bessel solutions (\ref{bessol})
or even an approximation to this.
The characteristic delocation time interval in the above is
\begin{equation}
\label{tint}
t_{deloc}=\frac{m\sigma^2}{\hbar},
\end{equation}
usually the raw variance $\sigma$ is taken to be a hand chosen delocalization length;
however taking it to be $\ell$ given by (\ref{critlen}) gives the characteristic time
\begin{equation}
\label{tcrit}
\tau=\frac{\beta^4\hbar^3}{64G^2m^5}.
\end{equation}
\section{Conclusion.}\label{conc}
The critical lengths and times for typical masses are given in the table below:  
for elementary particles they are too long to be measured whereas for astronomical sized
particles they are too short to be observed,
in both cases any effect would be masked by other factors;
however for objects the size of a virus there is a possibility of a measurable effect.
The long range for elementary particles might suggest that
they have an effect in the 'next' universe,  see \cite{carr},  however relativistic cosmology
has the speed of light built into it from the beginning so that the critical length
(\ref{critlen}) is not really applicable.

The static generalization of (\ref{bessol}) to the Klein-Gordon equation is immediate
as the time derivative terms do not enter;   however in the time dependent case
generalization of (\ref{nstaticsol}) to the Klein-Gordon equation is unlikely
to have a similar form as (\ref{nstaticsol}) has single powers of $\hbar$ which do not
occur in the Klein-Gordon equation.   No method is known to generalize to the Dirac equation.

Comparison can be made between the values in the table and other known radii:
the classical electron radius is $r_e=e^2/(mc^2)=3\times10^{-15}m$
and the Bohr radius is $a_0=(4\pi\epsilon_0\hbar^2)/((m_ee^2)=5\times10^{-11}m$,
both of which are of orders of magnitude different from the values of $\ell$
in the table below.   
Note that the denominator of the Bohr radius is of similar form to (\ref{critlen})
when $e\rightarrow m$.
These radii govern lattice spacings and cross sections,  but it is not clear what if anything
is a lattice dependent on $\ell$ or how cross sections could depend on it;
presumably any lattice spacing would be about the size of a virus which is larger than usual.

Using the Compton wavelength $r_c=\hbar/(mc)$,
the Schwarzschild radius $r_s=(2Gm)/c^2$,
Planck units (\ref{planckunits}),
the critical length (\ref{critlen})
and the critical time (\ref{tcrit}),
it is possible to produce the dimensionless ratios
\begin{equation}
\label{ratios}
\frac{r_s}{r_c}=2\left(\frac{m}{m_p}\right)^2,~
\frac{r_c}{\ell}=\frac{8}{\beta^2}\left(\frac{m}{m_p}\right)^2,~
\frac{r_s}{\ell}=\frac{16}{\beta^2}\left(\frac{m}{m_p}\right)^4,~
\frac{t_p}{\tau}=\frac{64}{\beta^4}\left(\frac{m}{m_p}\right)^5,
\end{equation}
which shows that the are no new dimensionless quantities except the universal mathematical
constant $\beta$,   the only dimensionless quantity
which takes a different value for each particle is the mass in Planck units.
\bc Table of Characteristic Quantities.\ec
\begin{eqnarray}
\label{table}
\begin{array}{llll}
{\rm Particle}            & {\rm Mass~ in~ Kg.} & {\rm Distance~ \ell~ in~ meters}  & {\rm Time~ \tau~ in~ seconds}\\
{\rm Electron}            & 9\times10^{-31}     & ~~~~~~4\times10^{32}              &  2\times10^{69}    \\
{\rm Proton}              & 2\times10^{-27}     & ~~~~~~3\times10^{22}              &  3\times10^{52}    \\
{\rm Lead~ atom}          & 4\times10^{-25}     & ~~~~~~5\times10^{15}              &  9\times10^{40}    \\
{\rm Buckyball~ molecule} & 1\times10^{-24}     & ~~~~~~3\times10^{14}              &  9\times10^{38}    \\
{\rm Protein}             & 6\times10^{-23}     & ~~~~~~1\times10^{9}               &  1\times10^{30}    \\
{\rm Haemoglobin}         & 1\times10^{-22}     & ~~~~~~3\times10^{8}               &  9\times10^{28}    \\
{\rm DNA}                 & 2\times10^{-21}     & ~~~~~~4\times10^{4}               &  3\times10^{22}    \\
{\rm Small~ virus}        & 7\times10^{-20}     & ~~~~~~9\times10^{-1}              &  5\times10^{14}    \\
{\rm Large~ virus}        & 1\times10^{-17}     & ~~~~~~3\times10^{-7}              &  9\times10^{3}     \\
{\rm Bacteria}            & 9\times10^{-16}     & ~~~~~~4\times10^{-13}             &  2\times10^{-6}    \\
{\rm Yeast}               & 6\times10^{-14}     & ~~~~~~4\times10^{-17}             &  3\times10^{-13}   \\
{\rm Man}                 & 9\times10^1         & ~~~~~~4\times10^{-64}             &  2\times10^{-91}   \\
{\rm Earth}               & 6\times10^{24}      & ~~~~~~1\times10^{-132}            &  1\times10^{-205}  \\
{\rm Sun}                 & 2\times10^{30}      & ~~~~~~4\times10^{-148}            &  3\times10^{-233}  \\
{\rm Galaxy}              & 6\times10^{42}      & ~~~~~~1\times10^{-186}            &  1\times10^{-295}  \\
{\rm Cluster}             & 1\times10^{46}      & ~~~~~~3\times10^{-196}            &  9\times10^{-312}  \\
\end{array}
\nonumber
\end{eqnarray}
\newpage
\section{Acknowledgement}
I would like to thank Tom Kibble for discussion on some of the topics in this paper.


\begin{thebibliography}{99}

\bibitem{barrow}
John D. Barrow (2002),
The Constants of Nature,
Random House,
ISBN9780099286479

\bibitem{carr}
Bernard J. Carr (2009),
Universe or Multiverse,
Cambridge University Press
ISBN-10:0521140692

\bibitem{cow}
R. Colella,  A.W.Overhauser,  S.A. Werner,
Observation of Gravitationally induced Quantum Interference.
{\it Phys.Rev.Lett.}{\bf 34}(1975)1472.

\bibitem{roberts}
Mark D. Roberts,
The Quantization of Geodesic Deviation.
{\it Gen.Rel.Grav.}
{\bf 28}(1996)1385-1392.

\bibitem{schiff}
Leonard I. Schiff (1968),
Quantum Mechanics,  Third Edition,
McGraw-Hill
ISBN 0-07-Y85643-5

\end{thebibliography}
\end{document}